\documentclass[useAMS,usenatbib, usegraphicx]{mn2e}

\newcommand{\npix}{A}

\newcommand{\SNTarget}{(S/N)_{\rm T}}
\newcommand{\SN}{S/N}

\newcommand{\Vbin}{\mathcal{V}}
\newcommand{\sumbin}{\sum_{k \in \Vbin_i}}

\newcommand\apj{ApJ}
\newcommand\apjl{ApJ}
\newcommand\mnras{MNRAS}

\title[Adaptive Binning with Weighted Voronoi Tesselations]{Adaptive
Binning of X-ray data with Weighted Voronoi Tesselations}
\author[S. Diehl \& T.S. Statler]{
Steven Diehl\thanks{E-mail:diehl@phy.ohiou.edu} 
and Thomas S. Statler\thanks{E-mail:statler@ohio.edu}\\
Astrophysical Institute, Department of Physics and Astronomy,
Ohio University, OH 45701}

\begin{document}

\date{Submitted soon.}
\pagerange{\pageref{firstpage}--\pageref{lastpage}} \pubyear{2005}

\maketitle
\label{firstpage}

\begin{abstract}

We present a technique to adaptively bin sparse data using weighted
Voronoi tesselations (WVTs). WVT binning is a generalisation of
Cappellari \& Copin's (2001) Voronoi binning algorithm, developed for
integral field spectroscopy. WVT binning is applicable to many types
of data and creates unbiased binning structures with compact bins that
do not lead the eye. We apply the algorithm to simulated data, as well
as several X-ray data sets, to create adaptively binned intensity
images, hardness ratio maps and temperature maps with constant
signal-to-noise ratio per bin. We also illustrate the separation of
diffuse gas emission from contributions of unresolved point sources in
elliptical galaxies. We compare the performance of WVT binning with
other adaptive binning and adaptive smoothing techniques. We find that
the CIAO tool {\it csmooth} creates serious artefacts and advise
against its use to interpret diffuse X-ray emission.

\end{abstract}



\section{Introduction} \label{Introduction}

X-ray data are generally very sparse in nature. To deal with this
problem, astronomers are often forced to either bin or smooth their
data. The most commonly used techniques are simply binning to square
blocks of a fixed size or convolving with a fixed kernel. However, due
to the large dynamic range in many extended objects, ordinary binning
and smoothing techniques are never able to capture structure on large
scales without masking detail on smaller scales. This deficiency is
the motivation for spatially adaptive algorithms.

With the advent of the two major X-ray satellites, {\it Chandra} and
{\it XMM-Newton}, it is now possible to resolve fine morphological
structures in extended X-ray emitting sources, such as galaxies,
clusters, or supernova remnants.  This calls for new techniques to
reliably extract spatial information. \citet[][hereafter
SF01]{Sanders} were the first to answer with a 2-dimensional adaptive
binning algorithm, applicable to background-corrected intensity images
and hardness ratio maps.  However, this algorithm is restricted to a
limited set of bin sizes, which prevents it from being fully adaptive
and from adjusting its resolution so as to keep the signal-to-noise
ratio (S/N) constant. This creates jumps in S/N of a factor of $\sim
2$, which, along with its quadrilateral bin shapes, can lead the eye
and suggest structure that is not there.

Motivated by the different problem of analysing 2-dimensional optical
integral field spectroscopic data, \citet[][hereafter
CC03]{Cappellari} developed an innovative adaptive binning technique
using Voronoi tesselations. Their algorithm is able to smoothly adjust
the bin size to the local S/N requirements and does not impose a
Cartesian geometry on the image. Unfortunately, it can be used only
with strictly positive, Poissonian or optimally weighted data whose
S/N is guaranteed to add in quadrature. This prevents it from being
useful in even simple situations in X-ray astronomy, involving data
corrected for exposure map effects or background, or in creating
hardness ratio maps.

In this paper, we generalise CC03's Voronoi binning technique so that
it can be used with any type of data. The generalised algorithm makes
use of Weighted Voronoi Tesselations (WVT), and combines the virtues
of both CC03's and SF01's techniques. It is as robust as, and even
more versatile than, SF01's code, yet retains the advantage of CC03's
flexible bin sizes. The algorithm produces smoothly varying binning
structures that are geometrically unbiased and do not lead the eye.

In section \ref{ExistingAbinning} of this paper, we review the two
binning techniques of SF01 and CC03 in more detail, pointing out their
advantages and drawbacks. In \S\ref{WVTbinning}, we explain the
functionality of the generalised WVT binning technique, and compare
its performance to the two older algorithms in section
\ref{Performance}. Section \ref{Applications} then demonstrates the
utility of WVT binning in creating X-ray intensity images, hardness
ratio maps, and temperature maps, and in disentangling the diffuse gas
emission in elliptical galaxies from the contribution of unresolved
point sources. Finally, \S\ref{Asmoothcomparison} quantitatively
compares WVT binning to commonly used adaptive smoothing algorithms,
before commenting on the availability of the code in section
\ref{Availability} and ending with conclusions in \S\ref{Conclusions}.


\section{Existing Adaptive Binning Algorithms} \label{ExistingAbinning}

\subsection{Quadtree Binning} \label{Quadtree} 

The pivotal work on spatial binning of sparse X-ray data is that of
SF01, who produce surface brightness and colour maps for the analysis
of X-ray cluster images.  Their algorithm starts with the smallest
possible bin size of $1\times 1$ pixel and calculates the S/N for each
bin. Each bin with a S/N higher than the user supplied minimal value
$(S/N)_{\textrm{\tiny{min}}}$ is marked as binned\footnote{Sanders and
Fabian's criterion of ``maximal fractional error'' is equivalent to a
``minimal S/N'' threshold.}, its pixel members are removed from the
pixel list, and ignored for the rest of the binning process. In the
next iteration, the unbinned pixels are rebinned with square bins of
double the side length. The S/N of each bin is computed, and those
exceeding the threshold are marked. This process is repeated until
either all pixels are binned or the bin size exceeds the image
size. Thus, the bins are generally square, with areas of $4^n$ pixels,
except for regions at the transition between two binning levels.
There, some pixels may have already been binned on a previous level
and removed.  The resulting bin shapes can be rectangular, L-shaped,
or more complicated. Even non-contiguous bins are common.

Owing to its hierarchical structure, which resembles a quadratic tree
commonly used in N-body simulations, we refer to this method as
``quadtree'' binning. Although slightly different implementations are
conceivable, we take SF01's version as representative. In section
\ref{Comparison_QT}, we make a rigorous quantitative comparison
between this algorithm and WVT binning.

\subsection{Voronoi Binning} \label{Voronoi}
              
Motivated by the need to optimally bin integral-field spectroscopic
data, CC03 present a method to spatially bin two-dimensional images
using Voronoi Tesselations. The goal is again to obtain a uniform S/N
per bin over the image, while keeping each bin as compact as possible.

A Voronoi Tesselation (VT) is a partitioning of a region, defined by a
set of points called the generators. Each point in the region, or in
this case, each pixel in the image, is assigned to the generator to
which it is closest. As a consequence of this scheme, the boundary
between two adjacent bins is always the perpendicular bisector of the
connecting line between the two generators (Figure \ref{WVT}).

A subset of VTs, called Centroidal Voronoi Tesselations (CVTs), has
the additional property that the generators coincide with the
centroids of the bins. CVTs are meaningful when there is a density,
$\rho$, defined over the region to be binned, and the generators are
the density-weighted bin centroids. Since the centroids cannot be
calculated before the bins themselves are constructed, it is necessary
to construct a CVT by iteration. A helpful tool is the Lloyd algorithm
\citep{Lloyd}, which iteratively constructs CVTs with generators at
each iteration taken as the centroids from the previous step. The
Lloyd iterations have the desirable effect of moving generators into
regions of higher density, thereby creating smaller bins. For a
uniform density, this algorithm tends to create hexagonal lattice
structures \citep{Du}.

In binning an image, one generally works with a signal $S_k$ per
resolution element $k$ (``pixel'' from now on)\footnote{For integral
field spectroscopy, this is the averaged signal over a fixed
wavelength interval.} and the associated noise per pixel, $\sigma_k$.
One can compute the S/N of a bin $\Vbin_i$ as
\begin{equation}\label{addSN}
(S/N)_i =  \frac{\sum_{k \in \Vbin_i}{S_k}}{\sqrt{\sum_{k \in
	\Vbin_i}{\sigma_k^2}}}. 
\end{equation}
For pure Poisson statistics, or certain forms of optimal weighting,
the $(S/N)^2$ is additive (CC03). With this restriction, one can make
use of a property of the Lloyd algorithm known as Gersho's conjecture:
applying the Lloyd algorithm to the square of the density tends to
produce a configuration with equal mass per bin \citep{Gersho}. CC03
exploit this conjecture by applying the Lloyd algorithm to the
quantity $\left(S/N\right)^2$, thus producing a CVT with a constant
S/N per bin.

In order for the Lloyd algorithm to converge, a good initial set of
generators is necessary. CC03 solve this problem with a
``bin-accretion'' algorithm. Starting from the pixel with the highest
S/N in the input image, one grows a bin by accreting nearest neighbours
until the bin reaches a minimum S/N or violates an imposed
``roundness'' criterion. Then the next bin is started from the pixel
closest to the weighted centroid of all previously binned pixels. This
method is guaranteed to generate compact bins within the desired S/N
range. Bins that do not meet both of these criteria are marked as
``bad'' and their pixels are reassigned to the next closest bin. The
centroids of the resulting bins are then used as the initial set of
generators for the Lloyd algorithm.

By definition, a Voronoi tesselation can produce neither gaps in the
data nor non-contiguous bins. The tesselation adjusts to uneven
boundaries smoothly, and the CC03 algorithm generally converges to a
solution with small, spatially independent scatter around the target
S/N. The bins are usually very compact, but can get more elongated or
ragged close to the boundaries, or in regions with very strong S/N
gradients. The principal drawback of CC03's algorithm lies in its
applicability. The algorithm works only for data in which S/N adds in
quadrature, as the iterative part is based on Gersho's
conjecture. This precludes the possibility of applying the code to
background-corrected or exposure-corrected data, hardness ratios, or
other types of data where $(S/N)^2$ is not additive.


\section{Adaptive Binning with Weighted Voronoi Tesselations}\label{WVTbinning}

\subsection{Introduction to Weighted Voronoi Tesselations (WVT)}

As described above, a normal VT assigns each pixel $k$ to the
generator $z_i$ to which it is closest; i.e., one finds the bin to
which the pixel belongs by minimising its distance to the generator
$|x_k-z_i|$ over all bins $i$. In order to make this definition more
flexible, we use a generalisation known as a Weighted Voronoi
Tesselation \citep[e.g.,][]{Moller}. In a WVT, each bin $i$ has an
associated scale length $\delta_i$ in addition to its generator $z_i$,
and a pixel $k$ is assigned to the bin that minimises
$|x_k-z_i|/\delta_i$. One can picture $\delta_i$ as a factor that
stretches or compresses the metric inside the bin $i$. An intuitive
analogy is simultaneous crystal growth, with the generators
representing the seeds and the scale length representing the growth
rate \citep{Moller}. A WVT is completely described by its set of
generators and scale lengths and can therefore be stored very
efficiently.

\begin{figure}
\includegraphics[width=84mm]{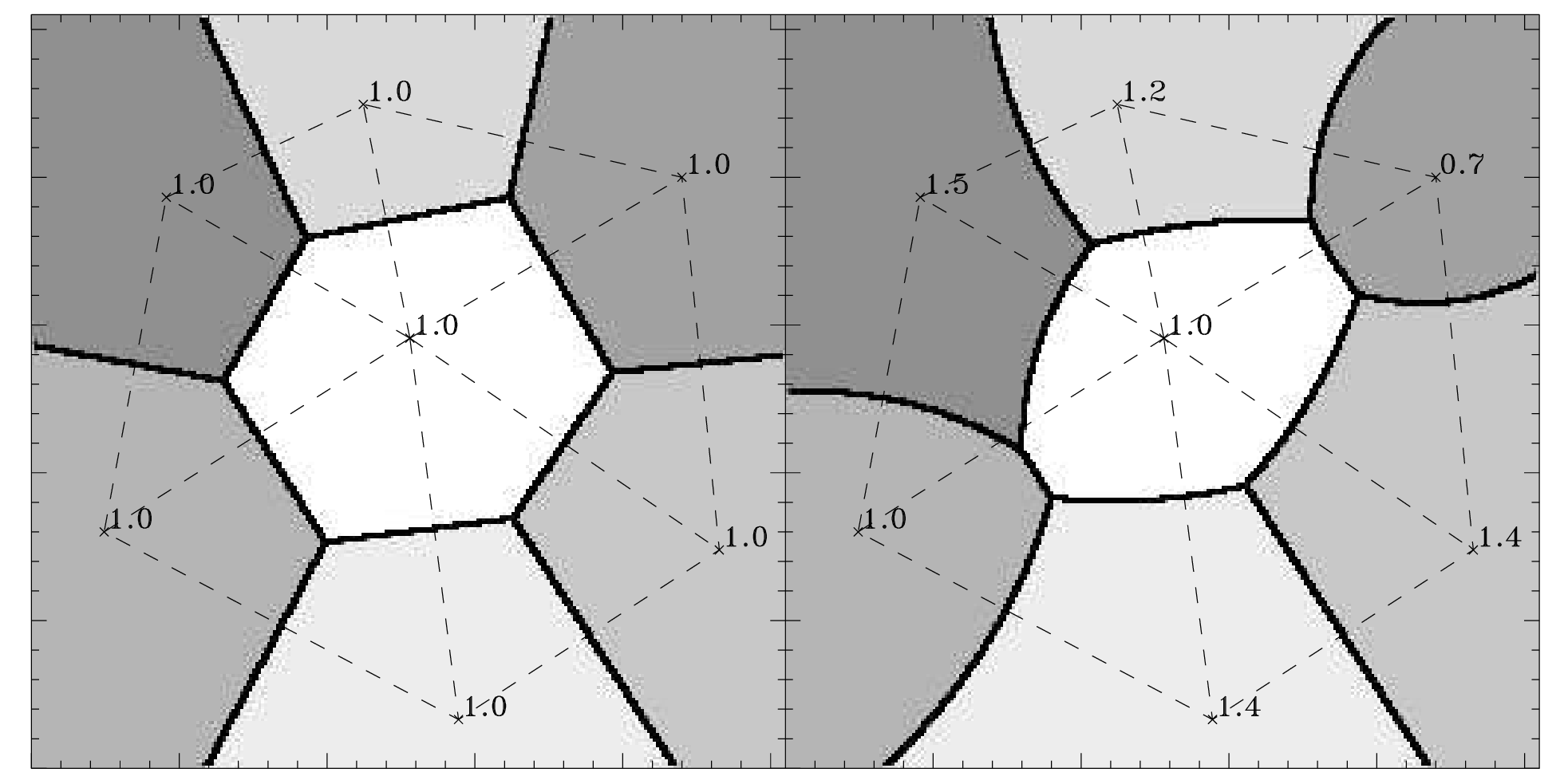}
\caption{A normal VT (left) and a WVT  (right) with identical bin
  generators $z_i$ (crosses). The numbers attached to the bins are the
  associated scale lengths $\delta_i$. The dashed lines connect
  neighbouring bin generators. Note how the bin boundaries are always
  perpendicular to them. For a normal VT, the bin boundaries
  are the perpendicular bisectors; for a WVT, the dashed lines
  are divided proportional to the respective scale lengths.\label{WVT}}
\end{figure}

Figure \ref{WVT} illustrates the appearance of a WVT (right) with a
simple example of bins with different relative scale lengths, and
compares it to an unweighted VT with the same generators (left). Note
how the bin boundaries move closer to the generators with the smaller
associated scale lengths, making their bins rounder and more
compact. For WVTs, the boundaries $b$ between two bins $i$ and $j$
always fulfill the equation $|b-z_i|/\delta_i = |b-z_j|/\delta_j$,
implying that the ratio of the bin radii is equal to the ratio of
scale lengths. This property is used below to modify the bin sizes by
manipulating their relative scale lengths and letting the generator
locations adjust. This replaces the Gersho prescription which changes
the bin sizes by explicitly moving the generators.

\subsection{Adaptive Binning Algorithm}

In the following discussion, we assume that there is some general way
to combine the signal and noise of various pixels to calculate the
resulting S/N for the bins. We emphasise that the details of how the
S/N is actually calculated are irrelevant for the functionality of the
WVT binning algorithm.

Our algorithm creates a weighted Voronoi tesselation, choosing the
scale lengths $\delta_i$ such that the bins have a near-uniform S/N
distribution, with the least possible scatter around the target
signal-to-noise $\SNTarget$. To find the appropriate scale lengths, it
is useful to consider the quantity $\mu_i$, defined by
\begin{equation}\label{mu}
\mu_i=\frac{(S/N)_i}{{\npix}_i},
\end{equation}
where $A_i$ is the bin area and $(S/N)_i$ is the S/N ratio in bin
$i$. The algorithm is aiming for a configuration where the bins have a
S/N equal to the target value, $\SNTarget$. In this configuration, we
should have, approximately,
\begin{equation}\label{mu2}
\mu_i = \frac{\SNTarget}{q\,\delta_i^2},
\end{equation}
where $q$ is a dimensionless constant that depends weakly on bin shape
(for circular bins, $q=\pi$). Combining equations (\ref{mu}) and
(\ref{mu2}) gives a rule for setting the scale length at each
iteration:
\begin{equation}\label{delta}
\delta_i = \sqrt{\frac{\SNTarget}{q\,\mu_i}} =
\sqrt{\frac{{\npix}_i}{q}\cdot\frac{\SNTarget}{(S/N)_i}}. 
\end{equation}
Since the binning depends only on the ratio of the scale lengths, the
value of $q$ is unimportant. We show below that good results are obtained
taking $q={\rm const}$ for all bins, regardless of shape. This
prescription replaces the Gersho-Lloyd procedure which, at each
iteration, moves the generators to the $(S/N)^2$ weighted
centroids. Such weighting is superfluous in our algorithm, so we
adopt the geometric bin centres as the new generators.

The WVT binning procedure thus proceeds as follows:
\begin{enumerate}
\renewcommand{\labelenumi}{(\roman{enumi})}
\item{Start with an initial WVT.}
\item{For each bin $i$, evaluate the signal to noise $(S/N)_i$, the
  area $A_i$ and the geometric centres $z_i$. \label{loopstart}}
\item{Calculate the scale length $\delta_i$ for each bin according to
  equation (\ref{delta}).}
\item{Reassign all pixels according to the new WVT with generators
  $z_i$ and scale lengths $\delta_i$.}
\item{Return to step (ii) until the bins stop changing significantly.}
\end{enumerate}
A binning with constant S/N across the field is a natural stable fixed
point of this iteration scheme, as it satisfies the relation
$\delta_i/\delta_j=\sqrt{A_i/A_j}$.

As in ordinary Voronoi binning, this algorithm requires a good set of
initial generators. We adopt CC03's solution of ``bin accretion'' with
a few modifications for speed, relaxing some acceptance criteria to
suit sparse and not strictly positive data (e.g. background subtracted
X-ray images). We also employ a soft lower S/N boundary for accepting
bins, in which the S/N has no longer to be larger than the fixed
target S/N. Instead, accretion terminates if the addition of another
pixel would increase the scatter around the target S/N. This
modification keeps the average S/N closer to the target value.

        
\section{Performance}\label{Performance}

\subsection{Comparison with Quadtree}\label{Comparison_QT}

While the simplicity of the quadtree binning algorithm makes it easy
to understand and apply, there are several disadvantages, which we
illustrate in this section using simulated X-ray data. A suitable
model for the surface brightness profiles $I(r)$ of a galaxy or galaxy
cluster is the circular $\beta$--model \citep{Sarazin_beta}:
\begin{equation}\label{betamodel}
I(r)=I_0\left[1+\left(\frac{r}{r_c}\right)^2\right]^{0.5-3\beta}+I_{Bg},
\end{equation}
where $I_0$ is the central surface brightness, $r_c$ is the core
radius, $\beta$ is a slope parameter, and $I_{Bg}$ is an additive,
flat background. We adopt the same parameters used by SF01: the image
size is set to $512\times 512$ pixels, $r_c$ is $128$ pixels, $I_0$ is
$100\,{\rm cts}\,{\rm pix}^{-1}$, $\beta$ is $0.67$ and $I_{Bg}$ is
$20\,{\rm cts}\,{\rm pix}^{-1}$. The simulated X-ray image is obtained
by populating the image with counts according to a Poisson
distribution. For the quadtree algorithm, a minimum S/N limit of $14$
($\sim 20/\sqrt{2}$) produces an average $\SN$ of $19.92$ in the test
image and is therefore chosen as the equivalent to a target S/N of 20
for the comparison with the WVT algorithm.

\begin{figure*}
\includegraphics[width=175mm]{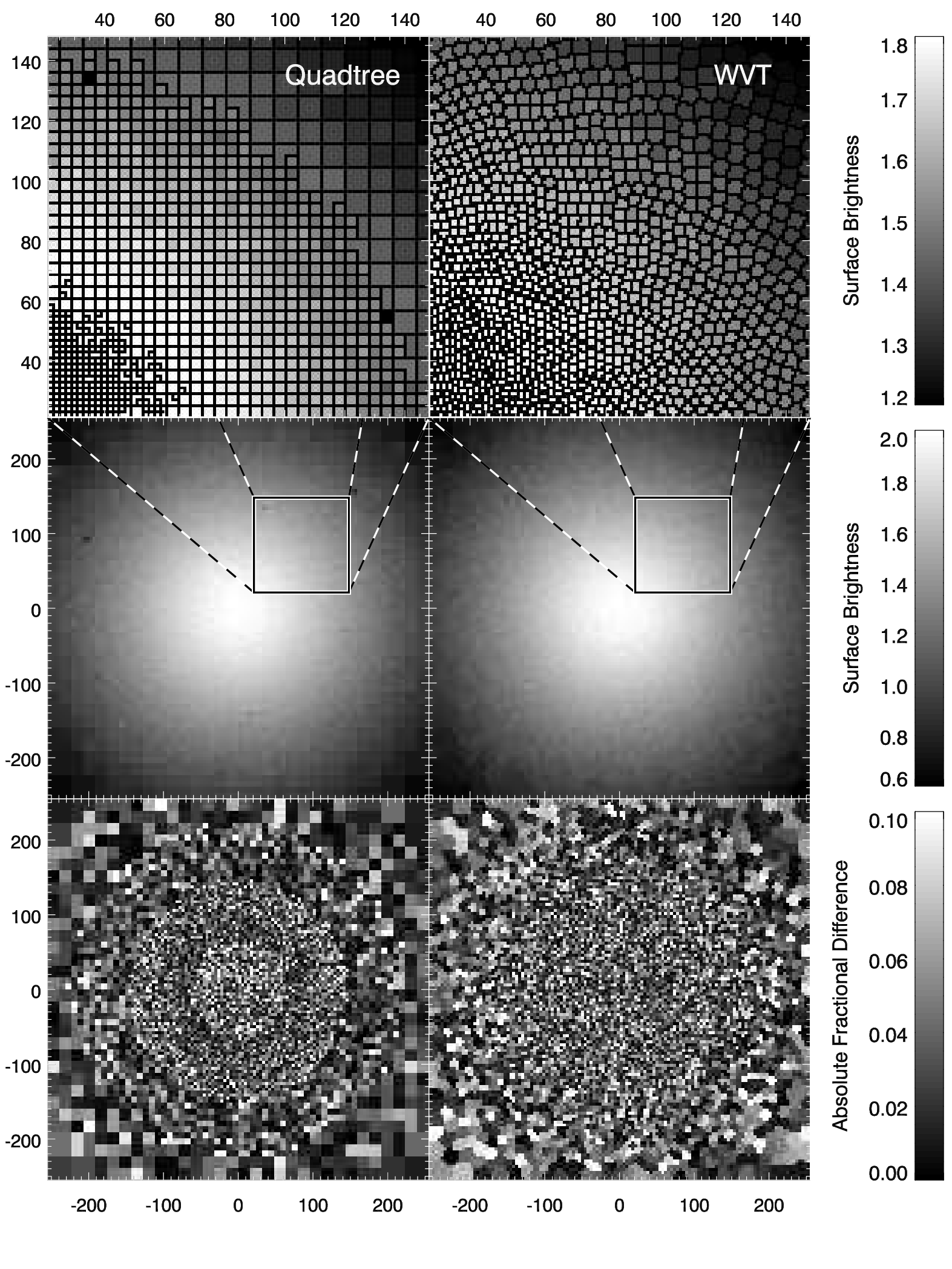}
\caption{Comparison between Quadtree (left column) and WVT binning
  (right column); {\em Middle Panels}: Logarithmically scaled, binned
  intensity images. The square indicates the region of the zoom-in
  shown in the {\em upper panels}. Each bin has been outlined to
  emphasise the difference in the binning structure. Note the darker
  ``stranded'' bins on the left; {\em Lower panels}: Absolute
  fractional difference between the model surface brightness and the
  binned simulated data.\label{fracdiff}}
\end{figure*}

The results of the quadtree and WVT binning algorithms are shown on
the left and right sides of Figure \ref{fracdiff}, respectively. The
middle panels show the full images; the upper panels zoom in on a
small region to emphasise the differences between the binning
structures. Note that the quadtree algorithm regularly forms
non-contiguous bins and can leave single pixels or small sets of
pixels ``stranded''. In a few cases, these pixels can be directly
picked out as isolated dark spots in the image, since a larger binning
level usually also corresponds to a lower average flux per bin. This
effect occurs predominantly in regions where neighbouring pixels have
already been binned on a previous binning level. SF01 offer two ways
to deal with this problem. The first is to handle isolated sets of
pixels of a non-contiguous bin separately, violating the minimum S/N
criterion as the bin is being split up. The alternative is to
redistribute the pixels to an adjacent neighbour bin. In the latter
case, the S/N of the neighbouring bin will be elevated, which can lead
to an increased scatter in S/N. At the same time, one sacrifices
resolution, as the effective number of bins is decreased. Throughout
the remainder of this discussion, we do not enforce contiguous bins in
the quad-tree algorithm for simplicity. We simply note that this
problem is absent in the WVT algorithm, which can easily be made to
enforce contiguous bins.

\begin{figure*}
\includegraphics[width=175mm]{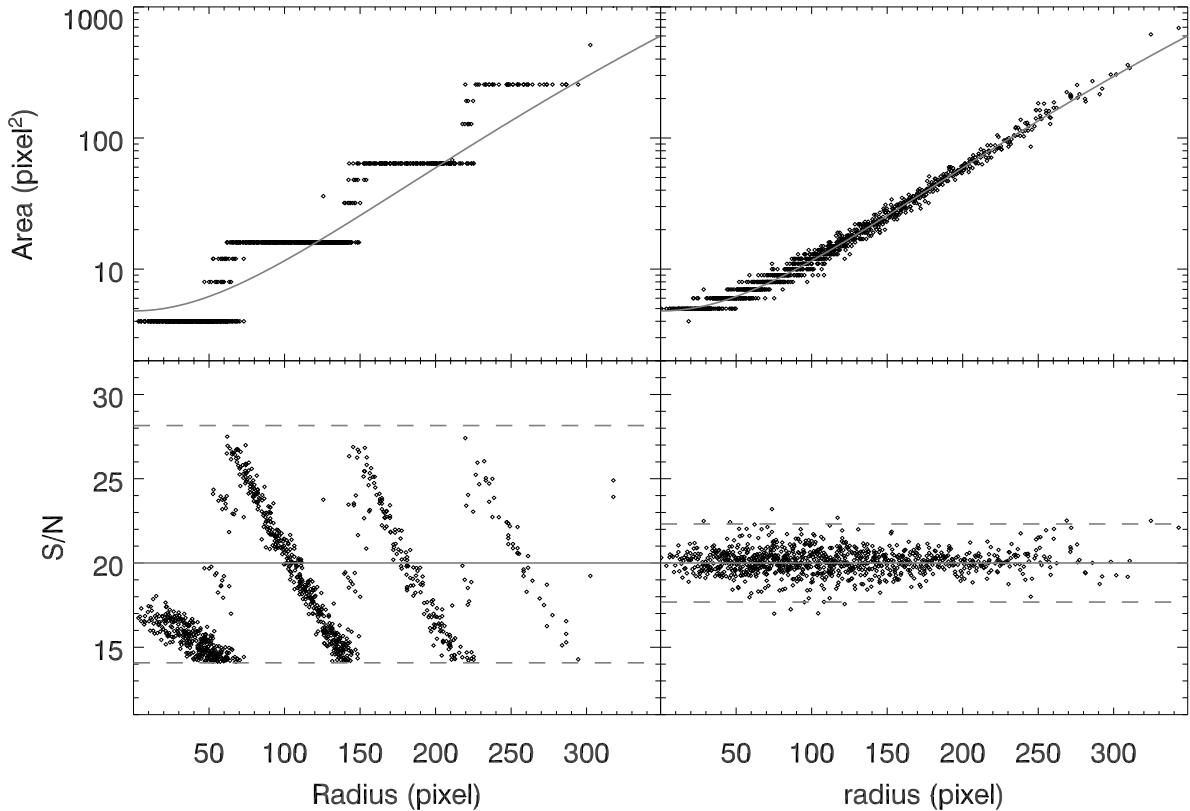}
\caption{{\em Upper panels}: area per bin vs. radius for quadtree
  ({\em left}) and WVT ({\em right}), the solid line indicates the
  theoretical prediction to produce a constant S/N of $20$ per bin for
  our test model.  {\em Lower panels}: corresponding S/N per bin; note
  the jumps in S/N due to the discrete bin sizes for the quadtree
  binning, which is completely absent in WVT; the solid line shows the
  target S/N, the dashed lines indicate the natural scatter of $\sim
  2$ in quadtree and the 3$\sigma $ rms scatter in
  WVT. \label{quadfig}}
\end{figure*}

The main problem with the quadtree algorithm lies in its small set of
discrete bin sizes. Except in small transition regions, where the bin
shapes are not square, the bin area is restricted to values of $4^n$
pixels. This discontinuous distribution of bin sizes is visible in the
binned image, and illustrated in the upper panels of Figure
\ref{quadfig}, which show the radial dependence of the bin areas for
quadtree (left) and WVT (right) binning. The solid line indicates the
optimal, theoretical bin size needed to produce the target S/N. The
discrete steps in the quadtree bin area translate into an
inhomogeneous S/N distribution, shown in the lower left panel of
Figure \ref{quadfig}. Each sharp increase in S/N corresponds to a
sudden jump in bin size, which decreases the local resolution beyond
the requirements of the target S/N. These jumps in S/N are easily
visible as circular rings in the fractional difference image in the
bottom left of Figure \ref{fracdiff}, showing that the binning
algorithm can create spurious structure.  In contrast, WVT binning
allows bins to adjust their size smoothly in single pixel steps, which
results in a reduced scatter around the target S/N and the removal of
misleading spurious features (bottom right panels of Figures
\ref{fracdiff} and \ref{quadfig}).

\begin{figure*}
\includegraphics[width=175mm]{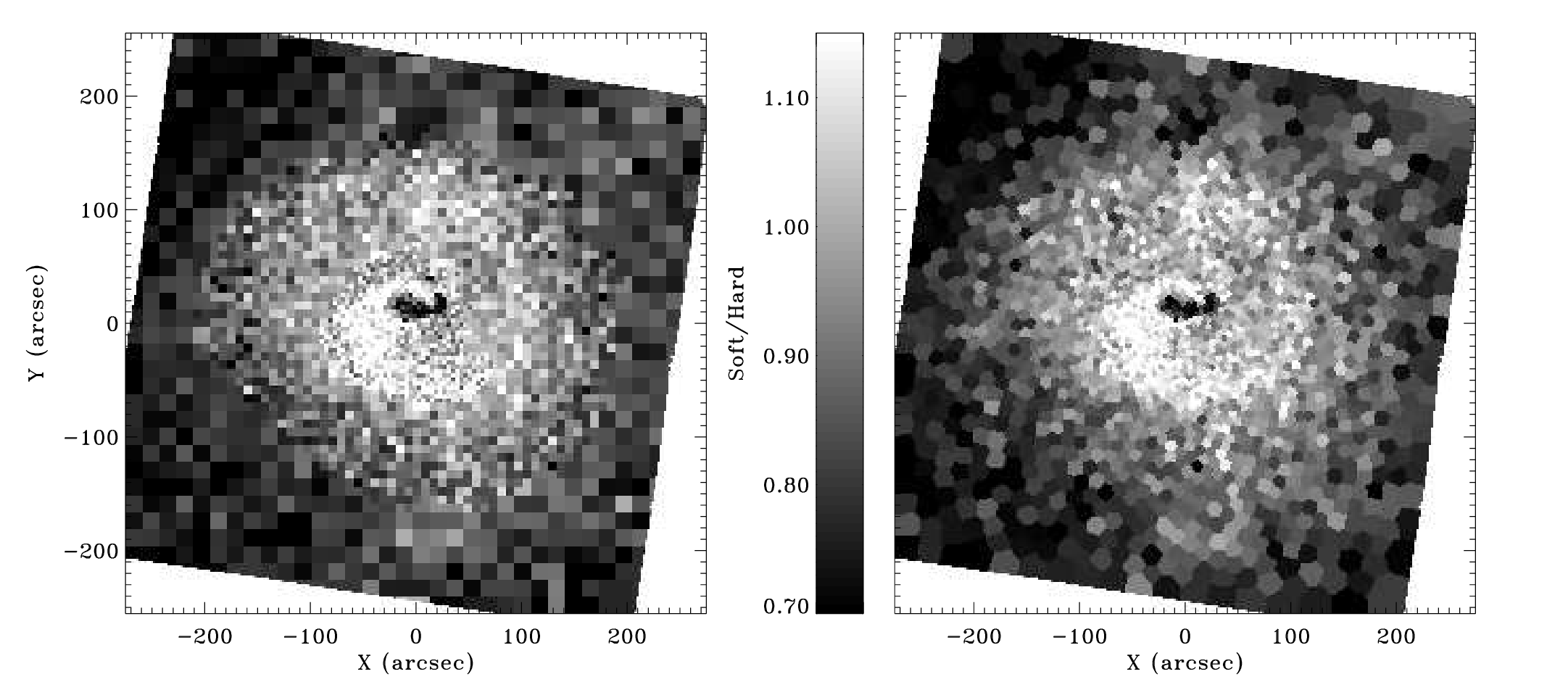}
\caption{Comparison of quadtree (left) and WVT binning (right); Both
  panels show adaptively binned hardness ratio maps of the core of the
  Perseus cluster, with dark colours indicating regions of higher
  temperature and/or lower photoelectric absorption. Note how quadtree
  binning leads the eye into identifying two ring structures, due to
  the strong jumps in the S/N where the bin area suddenly quadruples.
\label{color_Perseus}
}
\end{figure*}

The spatially correlated fractional error distribution resulting from
quadtree binning can be particularly misleading when the the bin value
is decoupled from the actual S/N distribution. An good example is a
hardness ratio map. Here, the signal is given by a flux ratio of two
independent bandpasses, whereas the S/N is determined by the total
flux of both bands. Figure \ref{color_Perseus} show quadtree and WVT
binned hardness ratio maps of the Perseus cluster. The eye identifies
two concentric rings in the quadtree binned map (left) at around 50
and 150 arcsec. These features are imprints of the discontinuous jumps
in bin size, and are completely absent in the WVT binned map
(right). WVT binned hardness ratio maps are described in more detail
in \S\ref{hardnessratiomaps}.

\subsection{Comparison with Voronoi binning}\label{Comparison_VT}

As the WVT algorithm generalises the method of CC03, which was
designed for optical integral field spectroscopic data, it is natural
to test the code on the same type of data. We apply our WVT binning
algorithm to the test data provided by Cappellari \& Copin, in their
on-line code release. The test data consist of a list of coordinates
and signal and noise values of the wavelength-integrated spectra of a
SAURON \citep{SAURON} observation of NGC 2273.

\begin{figure*}
\includegraphics[width=175mm]{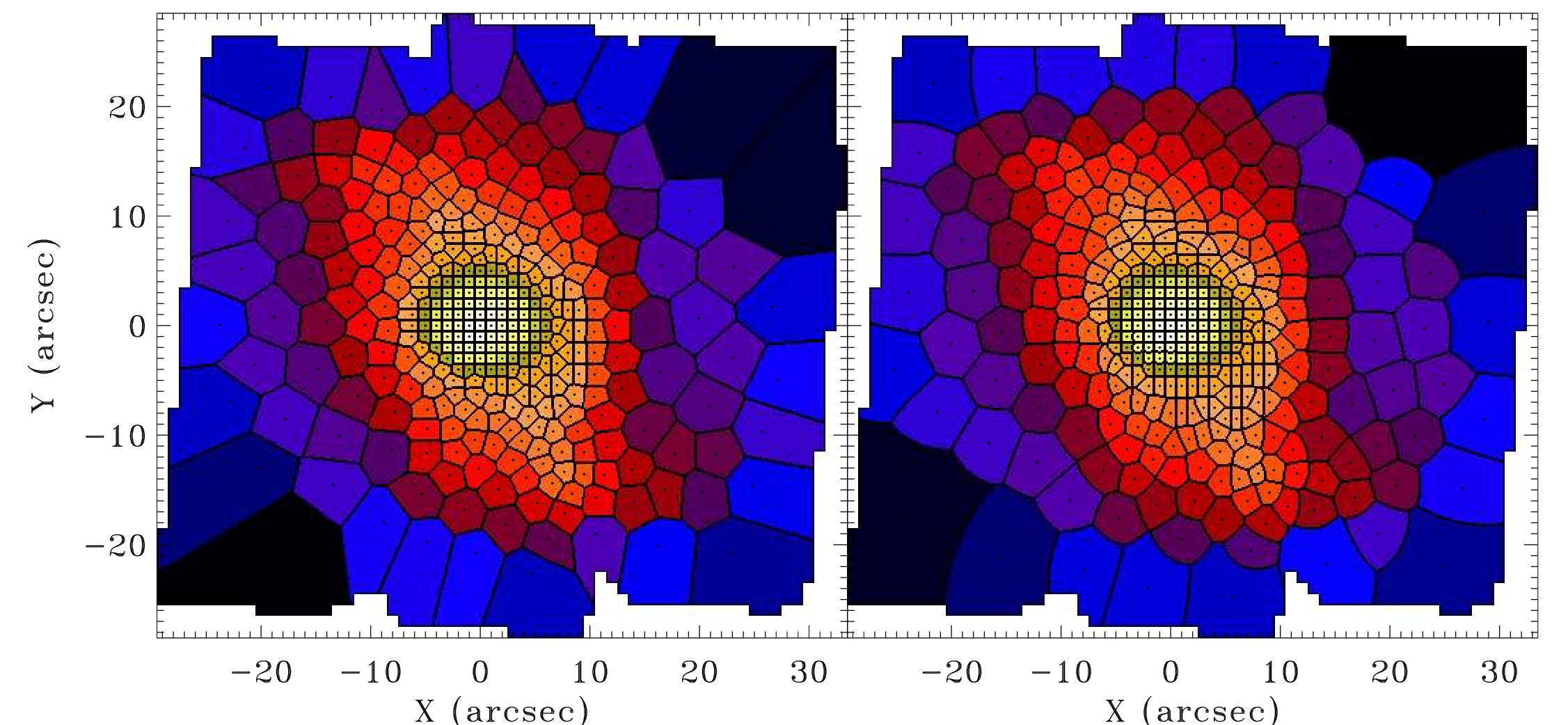}
\caption{Final result after application of the CC03's Voronoi binning
  code (left) and the WVT algorithm (right) to the SAURON data of NGC
  2273. After completion of the binning process, the bins have been
  projected onto a finer grid to make it easier to identify
  differences in the shape of the bins.\label{CappvsWVT}}
\end{figure*}

\begin{figure}
\includegraphics[width=84mm]{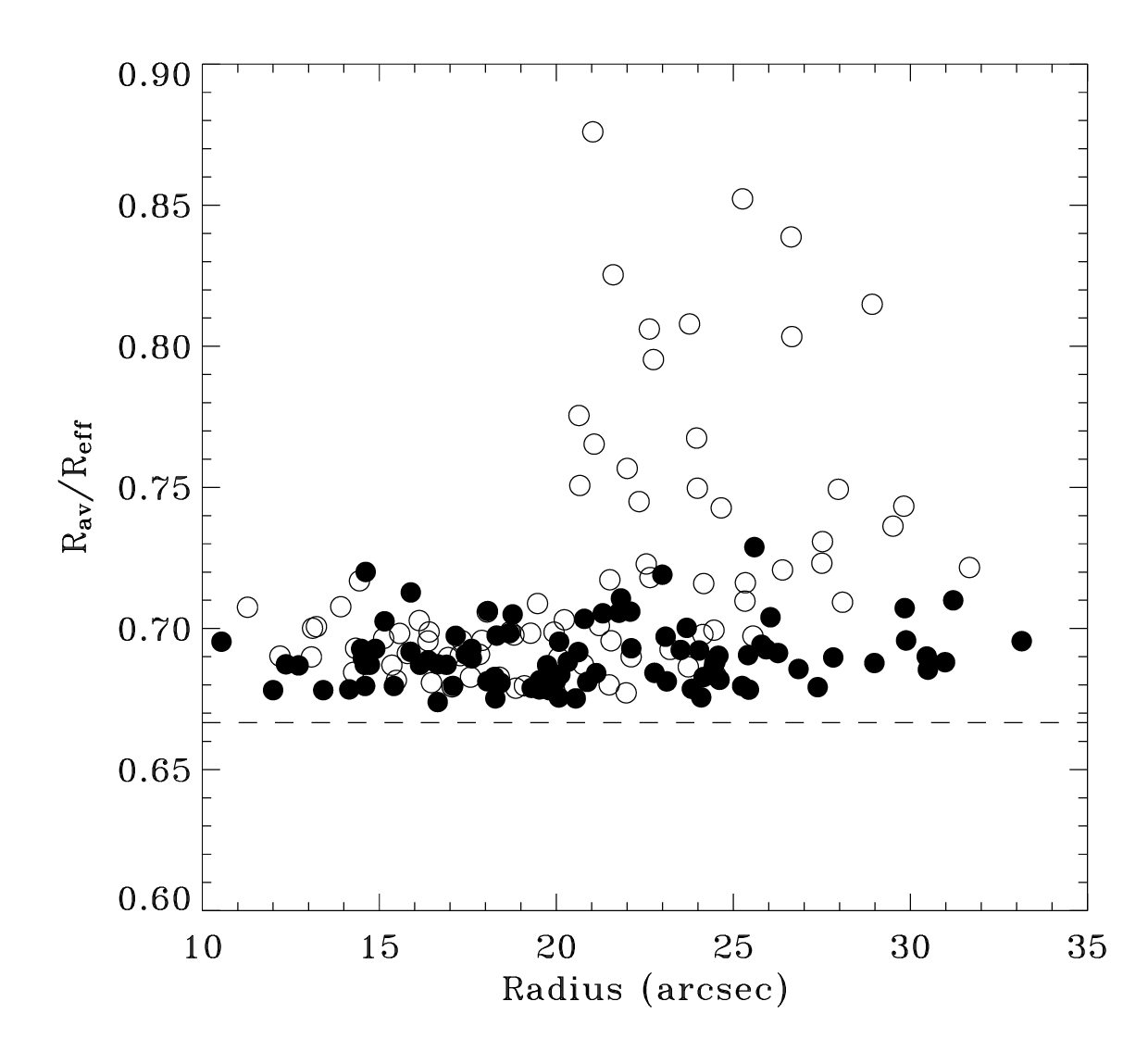}
\caption{Comparison of bin compactness between Voronoi binning (open
  circles) and WVT binning (filled circles). The dashed line indicates
  the theoretical limit for a perfect circle.\label{Roundness_CVT}}
\end{figure}

Figure \ref{CappvsWVT} compares the results of both algorithms for a
target S/N of $50$. Both yield consistent results with a comparable
scatter around the target S/N of only $\sim 6\%$. The only noticable
difference lies in the compactness of the individual bins, especially
close to the border and in regions of strong gradients. While CC03's
code tends to generate strongly elongated shapes in these cases, the
WVT bins stay consistently rounder. To quantitatively measure
roundness, we introduce the average bin radius $R^{\rm av}$ and the
effective bin radius $R^{\rm eff}$:
\begin{equation}
R^{\rm av}_i=\frac{1}{A_i}\sum_{j\in \Vbin_i} R_j,
\end{equation}
\begin{equation}
R^{\rm eff}_i=\sqrt{\frac{A_i}{\pi}}.
\end{equation}
The more compact the bin, the smaller is the ratio $R^{\rm av}/R^{\rm
eff}$. Its minimum at a value of $2/3$ represents a perfectly circular
bin. Figure \ref{Roundness_CVT} shows $R^{\rm av}/R^{\rm eff}$ as a
function of the distance from the galaxy centre and confirms that the
WVT algorithm (filled circles) produces more compact bins without edge
effects. At a distance of 20 arcsec, CC03's Voronoi bins (open
circles) get more elongated, as shown by the jump in
$R_{av}/R_{eff}$. This is mainly due to CC03's use of a weighted bin
centroid, which pushes the generators toward the bright end of the
bin, elongating the bins in the opposite direction.

\section{Applications to X-ray data}\label{Applications} 

\subsection{Intensity maps}

The most common application for adaptive binning in X-ray astronomy is
intensity binning. As discussed above, CC03's Voronoi binning
algorithm is valid only for purely Poissonian data. This would
correspond to raw counts for a perfectly flat detector response
without any background. However, real X-ray data are not as
simple. Many X-ray faint targets have surface brightness values
comparable to the background, which itself may be spatially
dependent. In addition, real observations exhibit strong variations in
effective area per pixel, due to chip gaps, node boundaries, or partly
overlapping multiple exposures. An observation's effective area $E_k$
per pixel $k$ is saved in an ``exposure map,'' which together with
the effective exposure time $\tau$, can be used to convert raw counts
$C_k$ per pixel into a flux $F_k$ with physical units 
\footnote{Alternatively, one can multiply the photon counts with their
detected energy to get units of ${\rm ergs}\, {\rm sec}^{-1} \, {\rm
cm}^{-2} \,{\rm arcsec}^{-2}$; see
http://cxc.harvard.edu/ciao/download/doc/expmap\_intro.ps for more
details on exposure maps.} of ${\rm photons}\, {\rm sec}^{-1} \, {\rm
cm}^{-2} \,{\rm arcsec}^{-2}$:
\begin{equation}
F_k=\frac{C_k}{E_k\,\tau}-B_k.
\end{equation}
Here, $B_k$ is the background flux per pixel. The variance in the same
pixel can be expressed as
\begin{equation}
\sigma_{F_k}^2=\frac{C_k}{E_k^2\,\tau^2}+\sigma_{B_k}^2,
\end{equation}
where $\sigma_{B_k}$ denotes the uncertainty that is attached to the
background value. The prescription for combining these quantities to
produce a S/N per bin is given by equation (\ref{addSN}). For the
hypothetical case where $\tau = 1$, $E_k=1$ and $F_{{\rm Bg}_k}=0$,
the signal $F_k$ reduces to pure counts and the binning scheme will
converge to a solution with a constant number of counts per bin.

\begin{figure*}
\includegraphics[width=175mm]{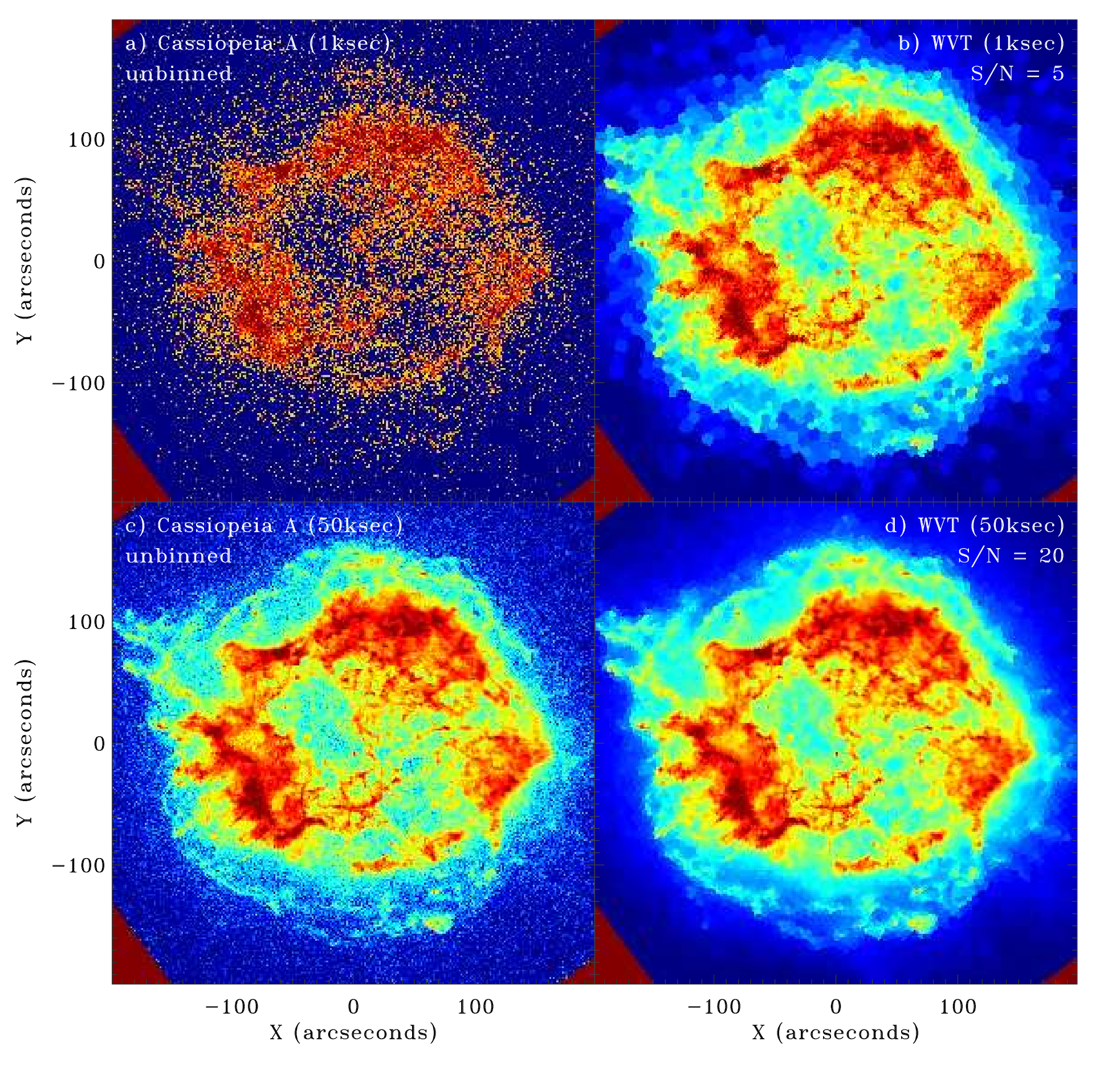}
\caption{ Top: Flux-calibrated {\it Chandra} image of Cassiopeia A
  with an exposure time of 1 ks (left) and the same data adaptively
  binned with WVT to a target S/N of 5 (right); Bottom: Cassiopeia A
  with the full exposure of 50 ks (left) and the same image binned to
  a S/N of $20$ (right).
\label{CasA}
}
\end{figure*}

We use the well-known 50 ks {\it Chandra} observation of Cassiopeia
A \citep{CasApaper} to demonstrate the power of adaptive binning for X-ray
images. The lower left panel of Figure \ref{CasA} shows the unbinned,
exposure map corrected counts image for the full exposure. In the
panel directly above, we restrict the data to only 1 ks of exposure
time. We bin this image with the WVT algorithm to a target S/N of $5$
(upper right). A comparison with the full 50 ks exposure shows that
WVT binning successfully reduces the noise, bringing out the
large-scale features in the outer parts of the image, while keeping
the appropriate resolution in the better exposed filamentary
features. Even in this short exposure, one is able to
pick out the central neutron star in the WVT binned image.
The image in the lower right of Figure \ref{CasA} shows the
full 50 ks image, adaptively binned to a S/N of $20$, to
demonstrate the applicability of WVT binning to a different S/N regime.

\subsection{Hardness Ratio Maps\label{hardnessratiomaps}}

Another useful tool in X-ray astronomy is the hardness ratio (or
``colour'') map. The hardness ratio $H^{AB}$ can generally be defined
as the quotient between the fluxes $F_A$ and $F_B$ in two different
bands $A$ and $B$, summed over all pixels of the bin $\Vbin_i$:
\begin{equation}
H^{AB}_i={\sumbin{F_{A,k}}\over{\sumbin{F_{B,k}}}}.
\end{equation}
The associated error can be expressed in terms of the noise in the
individual bands:
\begin{equation}
\sigma_{H^{AB}_i}=H_i\sqrt{{{(\sumbin{\sigma_{A,k}^2)}}\over{(\sumbin{F_{A,k}})^2}} +
{{(\sumbin{\sigma_{B,k}^2)}}\over{(\sumbin{F_{B,k}})^2}} }
\end{equation}
Depending on the choice of energy bands, a hardness ratio map can be
used as a diagnostic for any spectrally identifiable properties, such
as temperature gradients or photoelectric absorption features
(e.g. SF01). A general discussion about the physical interpretation of
these maps, an appropriate choice of bands, and a generalisation to
$n$ different bands can be found in SF01 or \citet{Fabian}.

We use the well-known 25 ks {\it Chandra} observation of the Perseus
cluster \citep{Fabian} to demonstrate a WVT binned hardness ratio
map. The right panel of Figure \ref{color_Perseus} shows a colour map
for the Perseus cluster, in which bright colours indicate regions of
lower temperature and/or lower photoelectric absorption. Our choice of
bands (A: $0.3-1.2$keV, B:$1.2-5$keV) shows both effects for
illustrative purposes; in principle, a different choice is able to
separate these two properties. The sharp dark feature close to the
centre is due to the photoelectric absorption ``shadow'' of an
infalling dwarf galaxy in the line of sight \citep{Fabian}. To the
north-east of the centre, one can pick out a giant radio cavity, with
cooler rims surrounding it. The smooth colour gradient toward the
centre also supports a cooling flow model \citep[see
e.g.][]{Sarazin_beta} and the ``swirl'' of the bright emission has
been interpreted as a sign for angular momentum of the infalling gas
\citep{Fabian}. Note that the WVT colour map does not contain the
spurious circular features present in its quadtree counterpart.

\subsection{Maps of Temperature (or other Spectral Parameters)}

Hardness ratio maps are often insufficient to disentangle the spectral
components of extended sources. This requires a detailed spectral
analysis of multiple regions within the field of view.  Current
state-of-the-art techniques to generate maps of temperature or other
spectral parameters usually specify a regular grid of points, within
which circular or square regions are ``grown'' until they reach a
minimum number of counts for the spectral analysis
\citep[e.g.][]{NulsenTempmap, OsullivanNGC4636}. One can extract a
spectrum and create response files for each region and feed them into
an X-ray spectral fitting package such as
Xspec\footnote{http://heasarc.gsfc.nasa.gov/docs/xanadu/xspec/},
ISIS\footnote{Interactive Spectral Interpretation System,
http://space.mit.edu/ASC/ISIS/} or
Sherpa\footnote{http://cxc.harvard.edu/sherpa/}. Just as in adaptive
smoothing (see also \S\ref{Asmoothcomparison}), the measurements in
this ``adaptively accreted'' temperature map are not independent of
each other. The user is burdened with the task of deciding which
features are actually resolved and thus believable, i.e. which
features are larger than the extraction regions. WVT binning, on the
other hand, has the power to automatically divide the field into
unbiased, independent bins with constant source counts per bin, while
keeping the individual bins as compact as possible. In addition, this
reduces automatically the number of time-consuming spectral fits from
the total number of pixels to the number of independent bins.

\begin{figure*}
\includegraphics[width=175mm]{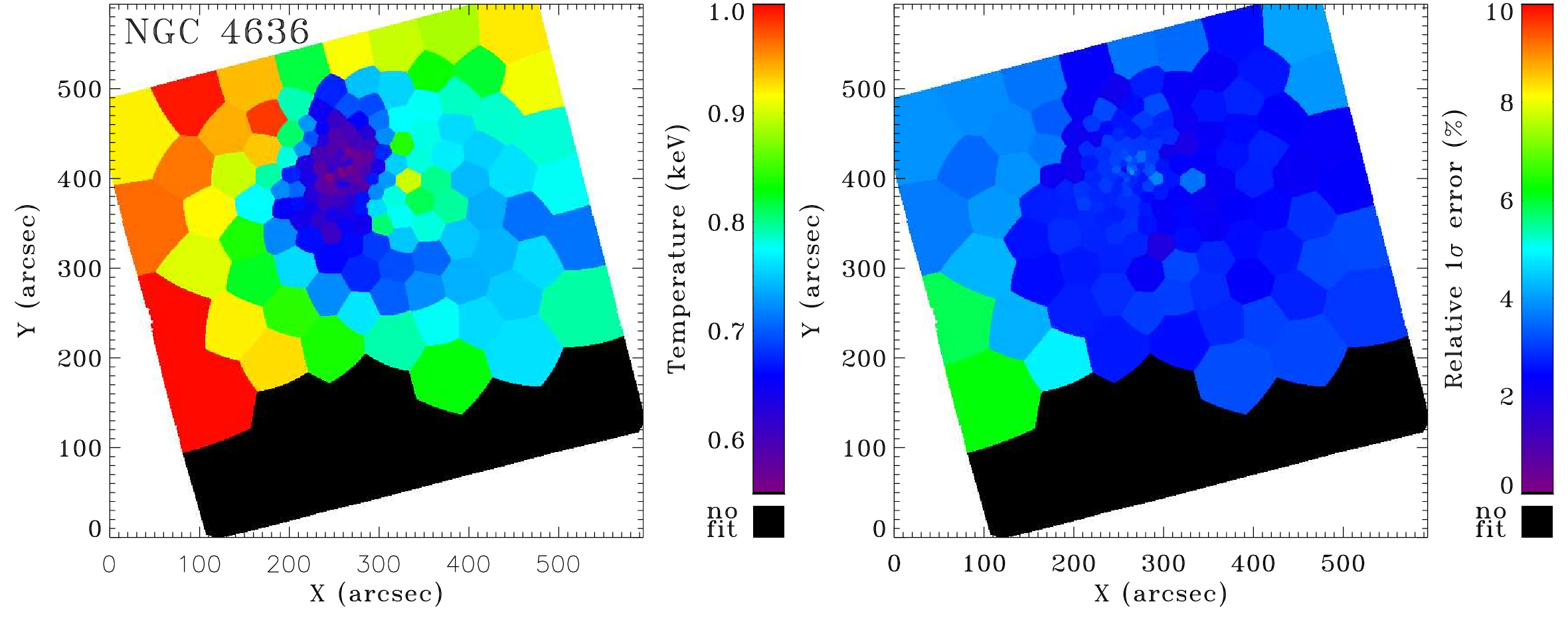}
\caption{Temperature map of NGC 4636. Temperatures are scaled linearly
  from $0.55$ to $1.0$keV, as indicated by the colour bar. The relative
  $1\sigma$ uncertainties, in percent, of the fitted values are shown
  in right panel. The error distribution is not completely uniform in
  this case owing to a combination of different levels in background
  contribution, degradation of the instrument response for large
  off-axis angles, and differences in the spectral shape for varying
  temperatures.\label{tempmap} }
\end{figure*}

To demonstrate this capability, we adaptively bin the 46 ks {\it
Chandra} observation of NGC 4636 \citep{JonesNGC4636} to 900 counts
per bin. We then extract a spectrum for each bin and fit it with an
absorbed, single temperature APEC\footnote{Astrophysical Plasma
Emission Code, http://cxc.harvard.edu/atomdb/sources\_apec.html}
model. The left panel of Figure \ref{tempmap} shows the resulting
temperature map, with the corresponding relative error distribution on
the right. This figure is directly comparable to Figure 2a of
\citet{OsullivanNGC4636}. They interpret the asymmetric temperature
distribution as the result of hotter gas surrounding the cool core of
NGC 4636, which is penetrated by a ``plume'' of gas extending to the
southwest. Inside of this plume sits a concave, rising radio bubble,
surrounded by cool rims. The higher temperature inside this cavity is
interpreted as a projection effect, as the bubble pushes the cooler
gas away and thus increases the relative contribution of the hotter
surrounding gas along the line of sight. The WVT binned temperature
map shows the same large-scale features as the
\citet{OsullivanNGC4636} map, with no ambiguity as to their
statistical significance.

\subsection{Isolation of different components}

In many astronomical sources, the observed emission comes from
multiple overlapping components. A good example are normal elliptical
galaxies, where the diffuse X-ray emission is made up of contributions
from interstellar gas and low mass X-ray binaries (LMXBs). Because of
their spectral differences, hot gas and LMXBs contribute differently
to the soft-band and the hard-band images. \citet{Diehl_XGFP,
Diehl_survey} show how this fact can be exploited to recover the gas
emission alone. Let $F_{S,k}$ and $F_{H,k}$ represent the
background-subtracted soft and hard images in each pixel $k$. We can
express both as linear combinations of the unresolved point source
emission $P_k$, the gas emission $G_k$, and their respective softness
ratios $\gamma$ and $\delta$:
\begin{eqnarray}\label{linearcombination}
  F_{S,k} &=& \gamma P_k + \delta G_k, \\
  F_{H,k} &=& (1-\gamma) P_k + (1-\delta) G_k. 
\end{eqnarray}
Thus, the uncontaminated gas image and its associated noise
can be expressed as 
\begin{equation}\label{linearcombtwo}
  G_k = {1-\gamma \over \delta-\gamma}\left( F_{S,k} - \left({\gamma \over 1-\gamma}
\right)\, F_{H,k}\right);
\end{equation}
\begin{equation}
\sigma_{G,k}={1-\gamma \over \delta-\gamma}\,\sqrt{\sigma_{S,k}^2+\left({\gamma \over 1-\gamma}\right)^2\sigma_{H,k}^2}.
\end{equation}
\citet{Diehl_XGFP} discuss the determination of 
the constants
$\gamma$ and $\delta$ with spectral models.

\begin{figure*}
\includegraphics[width=175mm]{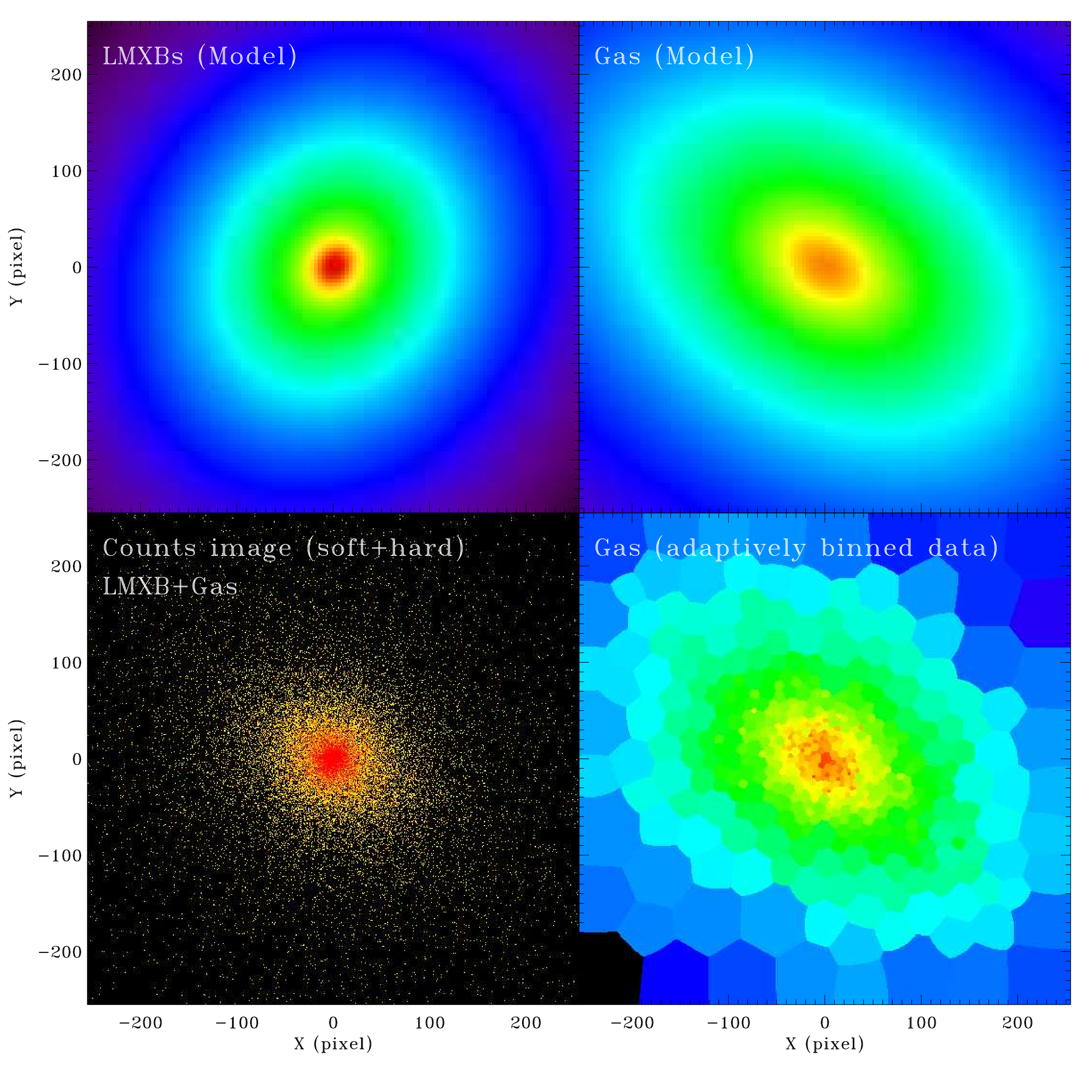}
\caption{Upper left: Model surface brightness distribution for
  unresolved point sources; Upper right: Model surface brightness
  distribution for the hot, isothermal gas; Lower left: Simulated
  Poisson image for the full band, including contributions of both
  point sources and gas; Lower right: Reconstruction of the
  gas surface brightness distribution with the help of WVT binning.
\label{gasimage} } 
\end{figure*}

Figure \ref{gasimage} demonstrates the isolation of the gas emission
using WVT binning, in a simulated observation. We assume gas and LMXB
sources with very different spatial distributions for purposes of
illustration. We adopt an elliptical de Vaucouleurs profile for the
LMXBs (top left), and a $\beta$ model, with a nearly orthogonal major
axis, for the gas (top right) and simulate hard and soft band images.
The bottom left panel of Figure \ref{gasimage} shows the full band
emission, which is nearly round.  Applying WVT adaptive binning (lower
right) to the gas image (equation [\ref{linearcombtwo}]), we are able
to reconstruct the true shape of the diffuse gas emission very
accurately.

\section{Adaptive Binning vs. Adaptive Smoothing}\label{Asmoothcomparison}

\subsection{Adaptive smoothing in X-ray astronomy}

Two adaptive smoothing algorithms are in widespread use due to their
inclusion in the main data analysis systems of {\it Chandra} and {\it
XMM-Newton}. The adaptive smoothing tool of the {\it XMM} Science
Analysis System (XMMSAS) is named {\it asmooth}, whereas the {\it
Chandra} Interactive Analysis of Observations (CIAO) tool is usually
referred to as {\it csmooth} (Ebeling, White \& Rangarajan, private
communication)\footnote{Note that the original {\it csmooth} code by
Ebeling, White \& Rangarajan was also named {\it asmooth}, although
the algorithm differs significantly from the XMMSAS tool. Whenever we
refer to {\it asmooth}, we mean the XMMSAS tool.}. Although their
output is generally not used for quantitative analyses, they have
become the primary tools to create ``pretty pictures'' for papers,
talks and press releases. Adaptive smoothing algorithms are thus
instrumental in forming and influencing the perceptions of the broader
astronomical community and the public.

In adaptive smoothing, the size of the smoothing kernel changes over
the field of view to create a constant S/N per pixel in the output
image.  It is worth emphasising that the number of independent
measurements is equal in adaptively smoothed and adaptively binned
images of the same target S/N. Thus one does not gain any additional
spatial information by smoothing rather than binning.

\begin{figure*}
\includegraphics[width=175mm]{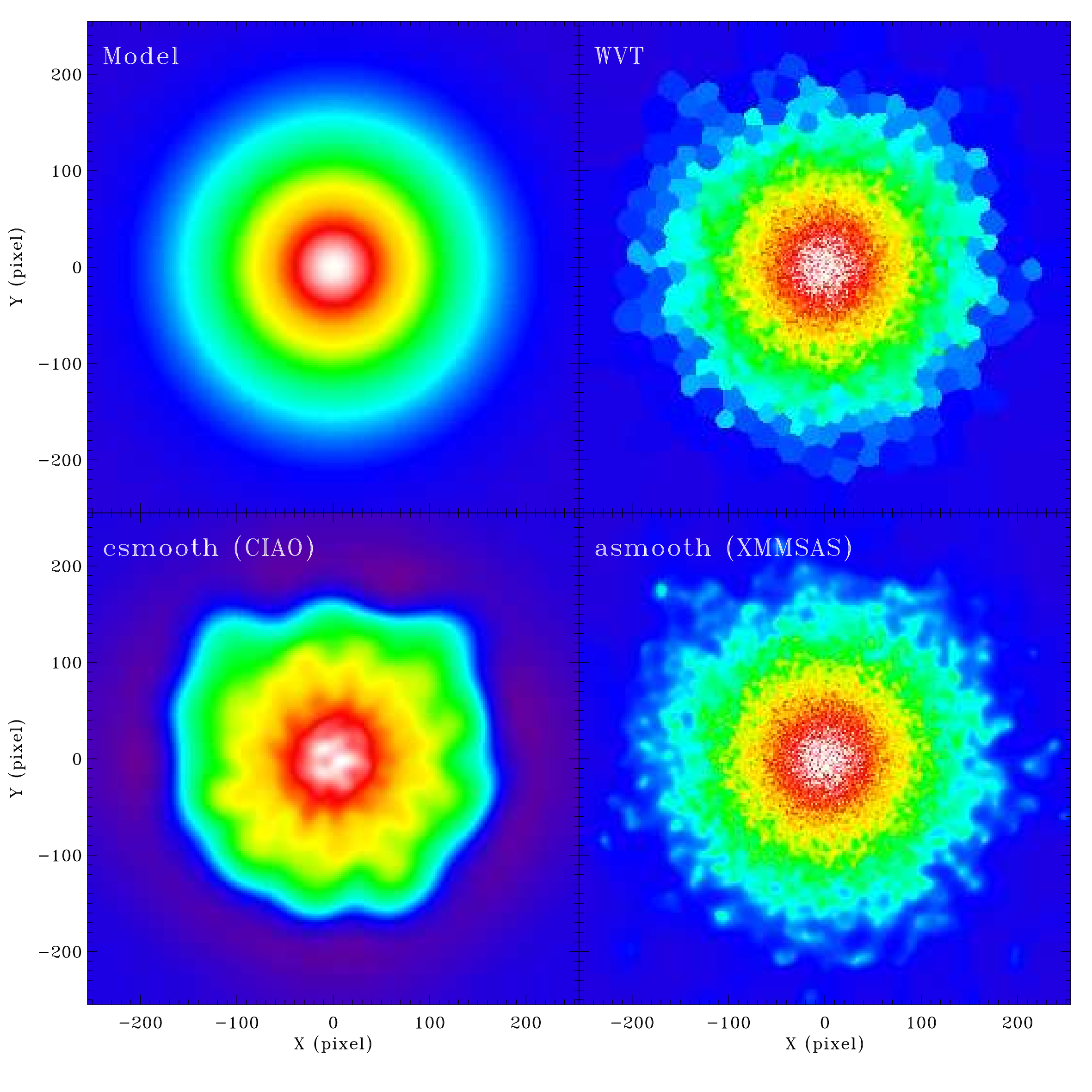}
\caption{A comparison of WVT binning with adaptive smoothing. Upper
  left: Model surface brightness distribution; The other three panels
  show the simulated counts image, adaptively binned image with WVT
  binning (upper right), adaptively smoothed with with {\it csmooth}
  (lower left) and {\it asmooth} (lower right). In the {\it csmoothed}
  image, note the radial ``fingers'', the annulus of deficient
  emission (deep purple) and the boundary effects in the corners.
\label{candasmooth}}
\end{figure*}

In this section, we give some cautionary advice on the interpretation
of adaptively smoothed images. We take a $\beta$--model surface
brightness distribution (equation [\ref{betamodel}]) with
$I_0=10\,{\rm cts}\,{\rm pix}^{-1}$, $\beta=0.67$, and $r_c=64\, {\rm
pix}$, with a background of $2\,{\rm cts}\,{\rm pix}^{-1}$, shown in
the upper left of Figure \ref{candasmooth}. We simulate a counts
image, which we adaptively smooth or bin to a target S/N of 5.

\subsection{Comparison with {\it asmooth}}

\begin{figure}
\includegraphics[width=84mm]{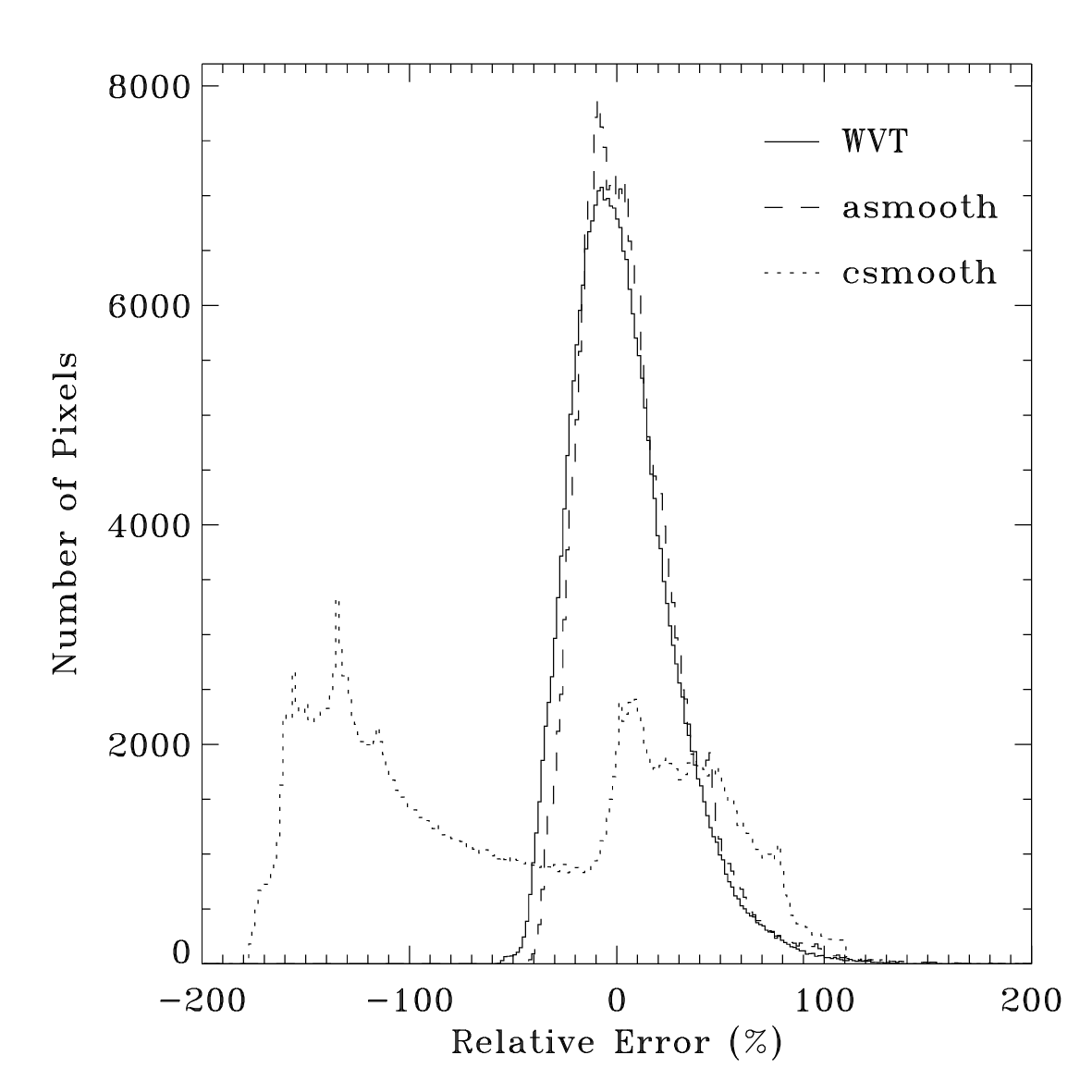} 
\caption{Histogram of relative errors, compared to the model surface
  brightness, for the example of Figure \ref{candasmooth}. The WVT
  binning (solid line) and {\it asmooth} (dashed line) histograms are
  consistent with the target S/N value of 5 (i.e. they approximate a
  Gaussian with a width of 20\%). Note that the adaptively smoothed
  image is {\it not} a statistically better representation of the true
  surface brightness. The histogram of {\it csmooth} results (dotted
  line) is very irregular with a wide range of positive and especially
  negative errors, demonstrating the failure of this algorithm.
\label{asmootherrhist}}
\end{figure}

The {\it asmooth} algorithm is thoroughly described in the
XMMSAS 6.0.0 user manual.
\footnote{http://xmm.vilspa.esa.es/sas/current/doc/asmooth/index.html}
The basic idea is to increase the size of the smoothing kernel for
each pixel until the pixel can ``accrete'' enough signal to meet the
S/N requirement. Thus, each pixel has a scale associated with it,
which determines the size of the convolution kernel that contributes
to the smoothed flux value at this point. In WVT binning, each {\it
bin} has a scale associated with it. In both cases, the scale is
determined from the local S/N distribution. Figure \ref{candasmooth}
shows the results of applying WVT binning (upper right) and {\it
asmooth} (lower right) to the same test model. Both algorithms are
able to reproduce the underlying surface brightness
distribution. Figure \ref{asmootherrhist} compares the distributions
of relative errors for {\it asmooth} (long dashed line) and WVT
binning (solid line). Both distributions are consistent with the
constant targeted S/N value of $5$, but {\it asmooth}'s error
distribution is not as regular as WVT's and is skewed slightly toward
higher fluxes.

\begin{figure*}
\includegraphics[width=175mm]{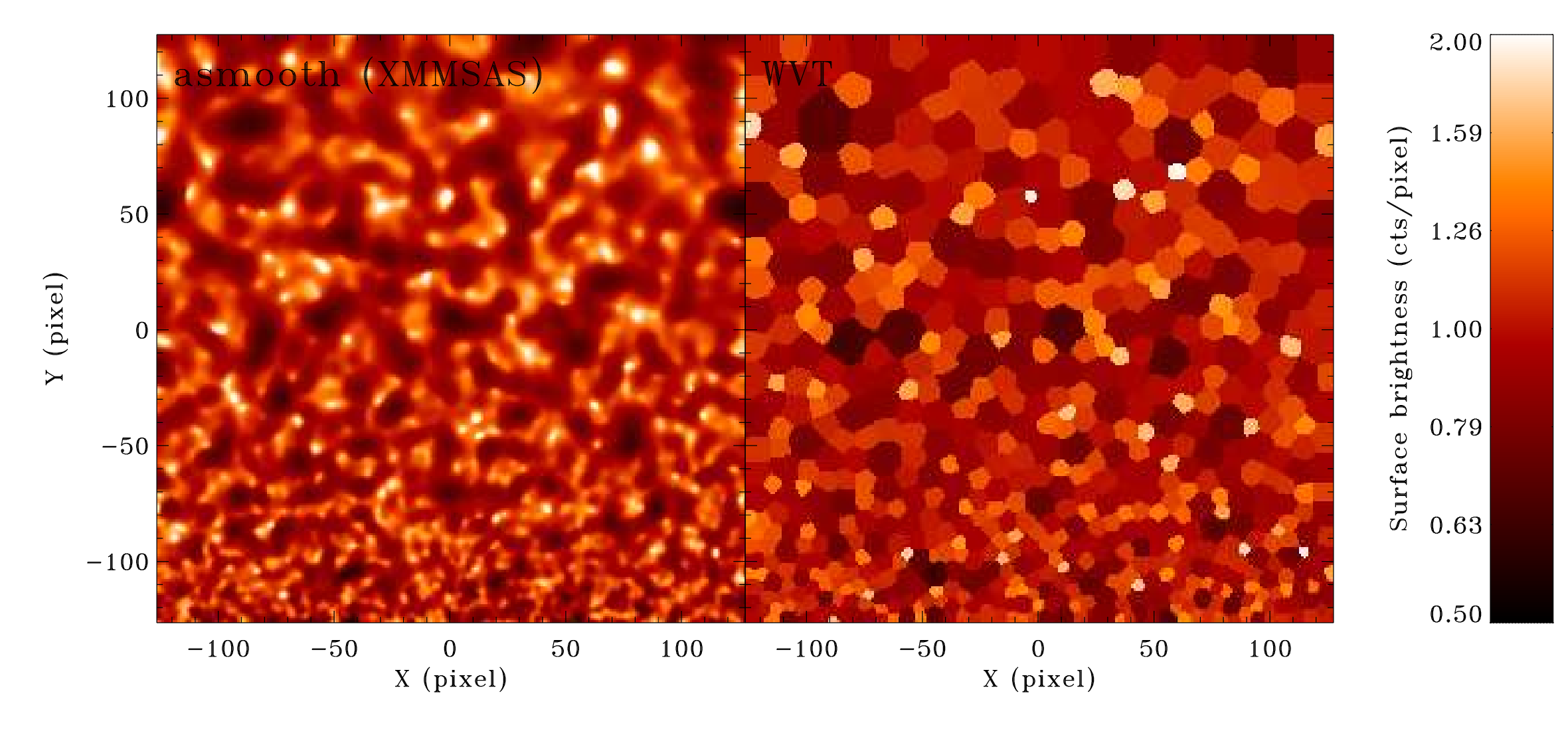}
\caption{Comparison between {\it asmooth} (left) and WVT binning
  (right): the simulated counts data was derived from a flat Poisson
  distribution with a count rate of $1\,{\rm cts}\,{\rm pix}^{-1}$,
  with a spatially variable background ``ramp'' increasing linearly from
  $0\,{\rm cts}\,{\rm pix}^{-1}$ at the bottom to $5\,{\rm cts}\,{\rm
  pix}^{-1}$ at the top of the image.
\label{flatbackground}}
\end{figure*}

The skewed error distribution is a result of {\it asmooth}'s tendency
to preferentially misidentify high flux pixels over low flux pixels as
real features. We find that the asmooth algorithm tends to build
``bridges'', connecting nearby, independent noise peaks, and making
them appear as linear filamentary structures. Examples can be seen in
the outer parts of Figure \ref{candasmooth}. A cleaner illustration is
shown in Figure \ref{flatbackground}. Here we have simulated a
flat-field image, with a vertical gradient in S/N. The figure compares
the results of {\it asmooth} (left) and WVT binning (right). The
smoothed image shows a wealth of spurious linear features that
strongly lead the eye, suggesting filaments and cavities. The binned
image, on the other hand, looks to the eye like a featureless but
noisy flat field. One can easily see that all of the apparent
structure in the smoothed image happens at the scale of the WVT bin
sizes, and is therefore not statistically significant despite the
target S/N of $5$ given as input.

In conclusion, if an adaptively smoothed image is necessary, we urge
that it be published only in conjunction with its smoothing scale map
or an equivalent WVT binned image.

\subsection{A cautionary note on CIAO's {\it csmooth}}

More than half of all {\it Chandra} press release images of the
diffuse emission from galaxies and clusters of galaxies are generated
with the CIAO tool {\it csmooth}. We demonstrate here that this
algorithm creates very serious artefacts and should be used only with
extreme caution.

The {\it csmooth} algorithm (Ebeling, White \& Rangarajan, private
communication) first calculates a set of smoothing kernel sizes,
ranging from the size of a single pixel to that of the entire
image. Starting with the smallest kernel, all pixels with a sufficient
S/N within the kernel to match the target S/N requirements\footnote{As
of CIAO 3.1, {\it csmooth}'s option to supply an external background
map to properly calculate the S/N distribution is not functioning
correctly. Instead, we use the default option to compute the
background from a local annulus surrounding the smoothing kernel. For
extended sources, this ``background'' region includes significant
amounts of the diffuse emission itself, resulting in strong S/N
variations across the field and a severe overestimate of required
smoothing scales.} are convolved and added to the output image. These
pixels are then removed from the input image, so they make no
contribution at larger scales. The algorithm then picks the next
larger kernel and starts over with the remaining pixels. This
continues until no more pixels are left in the image, or the kernel
size reaches its maximum. The final {\it csmoothed} image is the sum
of all these individually convolved slices, and the flux from each
pixel is spread over a different area, according to its smoothing
scale.

Unfortunately, there is a fundamental flaw in this algorithm. Because
a different smoothing kernel is assigned to each pixel in the input
image, each pixel in the output image is the sum of many convolutions
of different parts of the input image with different kernels. This is
fundamentally distinct from the {\it asmooth} algorithm, where the
kernel is assigned to the pixel in the output image, whose flux is
then a result of a single convolution using a single kernel. In other
words, {\it asmooth collects} flux, while {\it csmooth distributes}
flux. These procedures are identical only for pure convolution with a
fixed kernel. When the kernel is variable, the {\it csmooth} algorithm
has the effect of moving flux from low surface brightness regions into
high surface brightness regions. This is particularly destructive in
regions of relatively flat emission, where {\it csmooth} will move
flux into the high-flux tail of the noise distribution, creating
spurious emission features in the smoothed image. A good example is
given in the lower left panel of Figure \ref{candasmooth}, where {\it
csmooth} obviously produces spurious radial features.\footnote{This
image looks very similar to the claimed radial ``finger'' structures,
seen in the {\it csmoothed} image of NGC 4649 \citep{Randall}.} Note
also the annulus of depressed emission (deep purple colours) at a
radius of about 200 pixels. Here, the kernel reaches its maximum size,
dispersing the flux over a large area. The missing flux from this
annulus accumulates in regions with smaller smoothing scales,
producing a relatively sharp surface brightness edge around a radius
of 150 pixels. The relative errors in these two regions range from
+100\% to -200\% (Figure \ref{asmootherrhist}), indicating the
magnitude of this effect.


\section{Availability of the Code}\label{Availability}

WVT binning is implemented in IDL 5.6\footnote{http://www.rsinc.com},
and publicly available under http://www.phy.ohiou.edu/\~{}diehl/WVT. A
manual and download instructions are provided on the website, along
with examples for its usage.


\section{Conclusions}\label{Conclusions}

We have presented a generalisation of the Voronoi adaptive binning
technique by \citet{Cappellari}, broadly applicable to X-ray and other
data. The generalised algorithm exploits the properties of weighted
Voronoi tesselations, rather than the overly restrictive Gersho
conjecture.  WVT binning is applicable to any type of data as long as
there is a way to robustly calculate the S/N, and the S/N distribution
changes smoothly over the size of a bin.  We have demonstrated the
capabilities of WVT binning on exposure- and background-corrected
X-ray intensity images, colour and temperature maps, and in isolating
the diffuse gas emission in elliptical galaxies.
 
WVT binning overcomes the shortcomings of both Voronoi and quadtree
binning, the latter of which is in growing use in X-ray astronomy.
\citet{Sanders_contourbinning} have recently published results using a
``contour binning'' algorithm, in which the bin boundaries follow the
isophotes of an adaptively smoothed image. This creates very irregular
and elongated bins, which lead the eye and introduce a shell-like
appearance. We are unable to make a rigorous quantitative comparison
with this technique, as the details are still unpublished.  However,
we reemphasise that our WVT binning produces an unbiased distribution
of compact bins, and does not lead the eye.

We have also demonstrated the pitfalls of adaptive smoothing, and
regretfully advise against the use of the CIAO tool {\it csmooth} for
images of diffuse emission, as it creates very serious artefacts. If
an adaptive smoothing technique has to be used, we recommend the
XMMSAS tool {\it asmooth} instead. However we urge that adaptively
smoothed images be published only in conjunction with the smoothing
scale map or an equivalent WVT binned image to facilitate the
identification of real structures.

\section*{Acknowledgments}

We would like to thank Michele Cappellari and Yannick Copin for
extremely helpful discussions and for making their Voronoi binning
code publicly available. Our implementation of WVT binning is inspired
by and based on their original algorithm.


\label{lastpage}

\end{document}